\documentstyle[twocolumn,aps,epsf,floats]{revtex}
\begin{document} 

\twocolumn[\hsize\textwidth\columnwidth\hsize\csname %
@twocolumnfalse\endcsname

\title {Unconventional properties of superconducting cuprates}
\author{Andrey V. Chubukov$^{1}$ and Dirk K. Morr$^{2}$}
\address{$^{1}$
Department of Physics, University of Wisconsin, Madison, WI 53706}
\address{$^{2}$
University of Illinois at Urbana-Champaign, Loomis Laboratory of Physics, 1100 West Green Street, Urbana, Illinois 61801}
\date{\today}
\draft
\maketitle 
\begin{abstract}

We present an explanation of the unusual peak/dip/hump
features observed in photoemission experiments on Bi$2212$
 at $T \ll T_c$. We argue that these features arise from the
interaction of the fermionic quasi-particles with overdamped spin 
fluctuations.  We show that the strong spin-fermion interaction 
combined with the feedback effect
on the spin damping due to superconductivity yields
 a Fermi-liquid form of the fermionic spectral function
for $\omega < 2 \Delta$ where
$\Delta$ is the maximum value of the superconducting gap,
and a non-Fermi-liquid form for $\omega > 2 {\Delta}$. In the Fermi-liquid
regime, the spectral function $A({\bf k}_F,\omega)$ displays a 
quasiparticle peak at $\omega = {\Delta}$; in the non-Fermi-liquid 
regime it possesses a broad maximum (hump) at $\omega \gg {\Delta}$. 
In between the two regimes, the spectral function has a dip at 
$\omega \sim 2 {\Delta}$. 
We argue that our theory also explains
the tunneling data for the superconducting density of states.

\end{abstract}
\pacs{PACS numbers:71.10.Ca,74.20.Fg,74.25.-q}

]

\narrowtext

In recent years, the bulk of 
studies of cuprate superconductors was focused on
their unusual normal state properties. Less attention was paid to
the behavior of cuprates in the superconducting state. 
It was generally 
believed that the superconducting behavior, even in underdoped cuprates, is
rather conventional in the sense that most experiments can be explained
in the framework of the BCS-type theory for a $d-$wave superconductor.
Recently, however,
this belief has been challenged by photoemission experiments on Bi$2212$
materials \cite{norman,shennat}. These experiments demonstrated
that even in slightly overdoped cuprates, 
the spectral function $A({\bf k},\omega)$ at $T \ll T_c$ and in the momentum region
near $(0,\pi)$ where the $d-$wave gap is at maximum is qualitatively
different from the one expected for a conventional superconductor. 
Specifically, in the conventional case, $A({\bf k},\omega)$
 possesses a single sharp
peak at $\omega = \sqrt{\Delta^2_{\bf k} + \epsilon^2_{\bf k}}$ 
where $\Delta_{\bf k}$ is
 the superconducting gap  and $\epsilon_{\bf k}$ is the fermionic dispersion.
 The photoemission data 
for Bi$2212$ do show a sharp quasiparticle peak 
near the Fermi surface, but they also reveal
two extra features in $A({\bf k},\omega)$: a dip at frequencies right
 above the peak and a broad maximum (hump) at somewhat larger 
frequencies.
Moreover, as one moves away from the Fermi surface,
the sharp peak
looses its intensity but does not disperse, while the position of the 
hump varies with ${\bf k}$ and gradually recovers the normal state dispersion. 

There have been several phenomenological conjectures in the
recent literature
that the sharp peak  measured in
photoemission below $T_c$ is related to the one seen in neutron scattering data
in $YBCO$ and is due to the appearance of a 
dispersionless mode of unknown origin below $T_c$ ~\cite{norman,girsh,zhang}.
  In the present communication, we
present an alternative explanation of the photoemission data. 
We argue that the unusual
superconducting properties of cuprates can be explained by  a
strong interaction between electrons and  overdamped spin 
fluctuations~\cite{spinfl}.
 Specifically, we show that the peak/dip/hump features in the
spectral function emerge  due to a combination of two effects: 
(i) an almost complete 
destruction of the Fermi-liquid behavior which  
eliminates the  quasiparticle peak in the normal state and gives rise 
to a hump in the spectral function at higher frequencies, 
and (ii) a reduction of the spin damping
at small frequencies in the superconducting state which,
 as a feedback effect, restores Fermi-liquid behavior of the 
spectral function in the frequency range $\omega < 2 \Delta$. 
As a result, the spectral function near the Fermi surface possesses
a quasiparticle peak at $\omega \sim \Delta$, a dip at 
$\omega \approx 2 \Delta$ where the spectral function experiences a crossover
to a non-Fermi liquid behavior,  and a hump
at a higher frequency. 
As ${\bf k}$ moves away from the Fermi surface, 
the hump disperses with ${\bf k}$ while the quasiparticle peak only looses 
its intensity as it 
cannot move farther in frequency than $2\Delta$. This behavior fully agrees
with the photoemission results~\cite{norman,shennat}.

\begin{figure} [t]
\begin{center}
\leavevmode
\epsfxsize=7.5cm
\epsffile{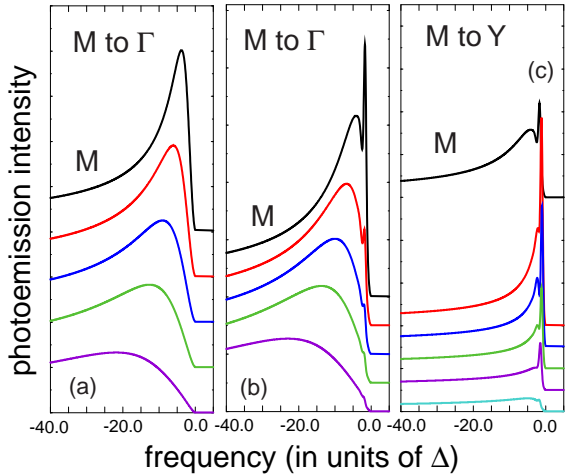}
\end{center}
\caption{The calculated quasiparticle spectral function 
in the normal state {\it (a)} and in the
superconducting state along $M -\Gamma$ ($(0,\pi) - (0,0)$) {\it (b)}
 and $M-Y$ ($(0,\pi) - (\pi,\pi)$) {\it (c)}. 
The results are presented for $b=5$.
For the $M$ point, we used ${\protect\bar \epsilon} = {\protect\Delta}$.}
\label{sftot}
\end{figure}

We first briefly review the results for the spectral function in the normal 
state and then discuss our calculations in the superconducting state.
One of us has recently shown~\cite{chubukov} that the fermionic self-energy due
to the spin-fermion interaction is almost independent of momentum and has 
the form
${\bar g}^2 \Sigma(\omega) =
{\bar g}^2 2\omega/(1 + \sqrt{1-i |\omega|/\omega_{sf}})$ where
${\bar g}^2 \propto \xi$ is a 
dimensionless coupling ($\xi$ is the magnetic correlation length) and 
$\omega_{sf} \propto  ({\bar g}^2 \xi)^{-1}$ is the typical relaxation 
frequency of overdamped spin fluctuations. 
It was further argued that 
in cuprates ${\bar g}^2$ is large even for overdoped materials so that 
the self-energy overshadows the bare frequency dependence in the fermionic
propagator,  i.e., 
$G_n ({\bf k},\omega) \approx Z/(\Sigma (\omega) - \epsilon_{\bf k})$ where
$Z = {\bar g}^{-2}$, $\epsilon_{\bf k} = Z {\bar \epsilon_{\bf k}}$, and
${\bar \epsilon_{\bf k}}$ is the bare fermionic dispersion. 
For $\omega < \omega_{sf}$, 
$\Sigma (\omega)$ has the Fermi liquid form 
$\Sigma (\omega) \approx \omega + i 
\omega|\omega|/(4 \omega_{sf})$ and accordingly, the spectral function 
$A({\bf k}, \omega)$ has a 
conventional peak at $\omega =\epsilon_{\bf k}$ though with a 
reduced residue $Z$.
 For $\omega > \omega_{sf}$, however, the system crosses
over to a region which is in the basin of attraction of
 the quantum critical point, $\xi =
\infty$. In this region, 
${\bar g}^2 \Sigma (\omega) \approx A e^{i\pi/4} |\omega|^{1/2} {\rm sgn}(\omega)$
where $A=2 {\bar g}^2 \omega^{1/2}_{sf}$ is independent of $\xi$.
As a result, instead of a sharp quasiparticle peak,
the spectral function possesses only a broad maximum at $\omega = 
\epsilon^2_{\bf k} /(4 \omega_{sf}) \gg \epsilon_{\bf k}$.
Various NMR experiments have shown that $\omega_{sf}$ is 
anomalously small even if the correlation length is comparable
to the interatomic spacing 
($\omega_{sf} \leq 15$ meV even at optimal doping~\cite{pines}) 
Since this $\omega_{sf}$ is smaller than the resolution of the photoemission
experiments, the experimentally measured $A({\bf k},\omega)$  only
displays a broad maximum even for ${\bf k} \approx {\bf k}_F$ (see Fig.~\ref{sftot}a).

Consider now the same system at $T << T_c$. We argue that there are
two key effects associated with the superconducting state. 
First, the quasiparticle Green's function is modified due to the 
fermionic pairing.
In the BCS approximation we have
$G^{-1}_{sc} ({\bf k},\omega) = G_n^{-1} ({\bf k}, \omega) +
\Delta_{\bf k}/Z)^2 G_n (-{\bf k}, -\omega)$ 
 where 
$(\Delta_{\bf k}/Z)^2$ is the strength of the $d$-wave pairing susceptibility. 
In a Fermi-gas ($Z=1$) $\Delta_k$ is the pairing gap. 
In our case, however, $Z \ll 1$, $G_n ({\bf k}, \omega) = Z/(\Sigma (\omega) -
\epsilon_k)$, and we obtain
\begin{equation}
G_{sc} ({\bf k},\omega) = Z \frac{\Sigma (\omega) + 
\epsilon_{\bf k} }{\Sigma^2 (\omega) -
(\Delta^2_{\bf k} + \epsilon^2_{\bf k})} \ .
\label{scg}
\end{equation}
For $\omega < \omega_{sf}$, 
this form of $G_{sc} ({\bf k}, \omega)$ again 
yields a conventional quasiparticle peak at $\omega = 
\sqrt{\Delta^2_{\bf k} + \epsilon^2_{\bf k}}$. 
For $\omega > \omega_{sf}$, however,
the spectral function takes the form 
\begin{equation}
A({\bf k}, \omega) \propto \sqrt{|\omega|} \frac{|\omega| + E_k + 
 \epsilon_{\bf k} \sqrt{|\omega|/(2 \omega_{sf})} 
{\rm sgn}(\omega)} {\omega^2 + E^2_k}
\label{sfsc}
\end{equation}
where $E_k = (\Delta^2_{\bf k} + \epsilon^2_{\bf k})/(4 \omega_{sf})$,
and thus shows the same
features as the one in the normal state: it possesses a broad maximum at 
$\omega = E_k$, but no quasiparticle peak.  

Clearly, the possibility to observe a quasiparticle peak in cuprates near 
${\bf k} = (0,\pi)$ where $\Delta_{{\bf k}}$ is near its maximum value 
$\Delta$ depends on the ratio $b= \Delta/\omega_{sf}$. 
If this ratio is small, then at frequencies comparable to $\Delta$, 
the system is in the Fermi liquid regime and $A({\bf k},\omega)$ 
displays a peak, while if  $b \geq 1$, then at $\omega \sim \Delta$, 
the system is already in the non-Fermi liquid regime where 
$A({\bf k},\omega)$ only exhibits a broad maximum.
An earlier computation of $b$  in the spin-fluctuation
approach~\cite{chubukov}
has shown that $b$ increases with decreasing 
doping and becomes larger than one
already for slightly overdoped cuprates.
At this stage, we therefore only obtain a hump in $A({\bf k},\omega)$.

However, there is a second effect related to superconductivity which 
gives rise to a
quasiparticle peak in the spectral function even if $b \gg 1$.
Indeed, in the above analysis we have so far assumed that $\omega_{sf}$
is independent of frequency.
This, however, is not true in the superconducting state
as the opening of the superconducting gap 
reduces the spin damping at low frequencies and hence increases 
$\omega_{sf}$ in the same frequency range. 
 From this perspective, the parameter $b$ we introduced before
is identical to $b_\infty \equiv {\Delta}/\omega_{sf} (\omega = \infty)$
where $\omega_{sf} (\omega = \infty)$ is the value of 
$\omega_{sf}$ in the normal
state, while the existence of a quasiparticle peak 
is actually determined by $b_{\omega} = 
{\Delta}/\omega_{sf} (\omega)$ at $\omega \sim \Delta$.

The spin relaxation frequency $\omega_{sf}$ 
 is inversely proportional to the damping of a spin fluctuation 
at ${\bf Q}=(\pi,\pi)$ due to its decay
into a particle-hole pair. To obtain $\omega_{sf} (\omega)$
 we hence have to evaluate
the imaginary part of the particle-hole bubble. The normal state 
calculations have
been reported previously~\cite{chubukov} and yielded a frequency independent 
$\omega_{sf} = (3/16) Z v_F \xi^{-1}$ ( $v_F$ is the Fermi velocity). It is
essential that this result is independent of $\Sigma$ and  is therefore 
the same as for free fermions. This universality is a general 
consequence of the fact that the normal state self energy
depends only on frequency~\cite{kadanof}. 
Vertex corrections do modify $\omega_{sf}$,
but these corrections are very small  
numerically and can be safely neglected~\cite{cmm}.

We now turn to the superconducting state. 
Here we have to (i) combine two fermionic bubbles
 made of normal and
anomalous Green's functions and  (ii)  reevaluate
the fermionic self-energy $\Sigma({\bf k}, \omega)$ using superconducting Green's
functions for intermediate fermions.
 Simple estimates show that for finite 
${\Delta}$,
$\omega_{sf}$ indeed acquires a frequency dependence and 
recovers the normal state value only at $\omega \gg {\Delta}$. 
Moreover, unlike  normal state $G_{n}^{-1} ({\bf k},\omega)$, 
superconducting 
$ G_{sc}^{-1} ({\bf k},\omega)$ {\it cannot} be written as 
the normal state Fermi-gas result plus a momentum-independent 
self-energy. As a result,  $\omega_{sf} (\omega)$ depends on 
the self-energy of the intermediate fermions
which by itself depends on $\omega_{sf}$. Consequently, 
the equation for $\omega_{sf} (\omega)$ becomes an integral equation.
Below we solve for $\omega_{sf} (\omega)$  assuming that 
$\Sigma (k, \omega)$ in the superconducting state
has the same form as in the normal state albeit with a frequency dependent
$\omega_{sf} (\omega)$. 
We checked by explicitly evaluating the lowest-order self-energy
diagram that the terms we omitted by using the normal state rather than
the superconducting Green's function change the results by less
than $5\%$ for all frequencies.

To obtain the equation for $b_\omega$, we
integrated over the fermionic momentum in the
bubbles and assembling normal and anomalous contributions to the spin damping. This
yields
\begin{equation}
b_\omega = \frac{b}{\omega}~Re \int_0^{\infty} d \Omega~ 
\frac{\Delta^2 -\Sigma(\Omega_+)\Sigma(\Omega_-) +  
D(\Omega_+)D(\Omega_-)}{D(\Omega_+)D(\Omega_-)}
\label{inteq}
\end{equation}
where, we remind, $b_\omega = \Delta/\omega_{sf} (\omega)$, 
$b = b_\infty =\Delta/\omega_{sf} (\infty)$. Also $\Omega_{\pm} = \Omega \pm 
\omega/2$, $\Sigma (\Omega) = 2 \Omega/(1 + \sqrt{1-i 
|\Omega|/\omega_{sf}(\Omega)})$,
 and  $D (\Omega) = \sqrt{\Sigma^2 (\Omega)-\Delta^2}$.

\begin{figure} [t]
\begin{center}
\leavevmode
\epsfxsize=7.5cm
\epsffile{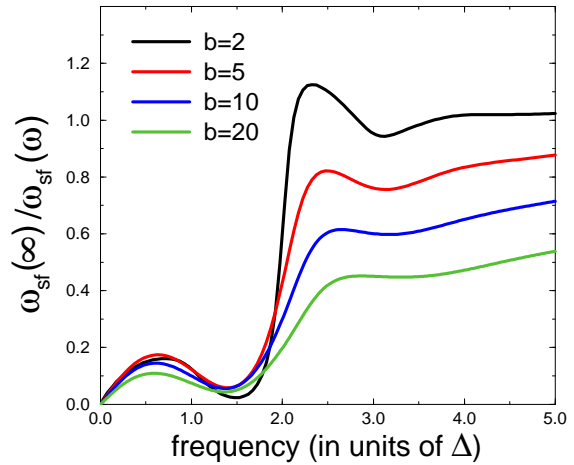}
\end{center}
\caption{The solution of the integral equation for the frequency dependent
spin-fluctuation frequency $\omega_{sf} (\omega)$ in
the superconducting state. In the normal state, $\omega_{sf} = \omega_{sf}
(\infty)$. For all $b =\Delta/\omega_{sf} (\infty)$, 
$\omega^{-1}_{sf} (\omega)$ sharply drops at around $2
{\protect\Delta}$.}
\label{omsf}
\end{figure}

It is instructive to consider the limits of small and large $b$. For 
$b \ll 1$, the fermionic damping is negligible, 
$\Sigma (\Omega) \approx \Omega$, and $b_\omega$ 
has the same functional form as the 
spin damping in the superconducting
Fermi gas: it is almost zero for 
$\omega < 2{\Delta}$, jumps to
$ b_\omega = b\pi/2$ at ${\omega} \geq 2\Delta$ 
and gradually
approaches $b$ with increasing frequency. 
A more careful look into Eq.(\ref{inteq}) shows that
the jump in $b_\omega$ at ${\omega}=2\Delta$ 
is a consequence of a  singularity in the integrand
at $|\Omega_{\pm}| = \Delta$.
For finite fermionic damping, the singularity is washed out but as long as
the damping at $|\Omega_{\pm}| = \Delta$ is small
 (which is clearly the case for
small $b$), $b_\omega$ 
sharply drops below ${\omega} =2\Delta$. 
Consider next
what happens when $b \gg 1$. For $\omega \gg \Delta$, 
$\omega/\omega_{sf} (\omega) \gg 1$ and hence $\Sigma (\Omega_\pm) \propto
e^{i\pi/4} \sqrt{|\Omega_\pm|}$.   
Using this form of $\Sigma (\Omega_\pm)$, one can easily
verify that the downturn renormalization of 
$b_\omega$  starts
already at $\omega \sim  \Delta b \gg \Delta$.
At smaller frequencies, we
expand the integrand in  (\ref{inteq})
in powers of ${\omega}$
and obtain $b_\omega \sim \omega/\Delta$.  
For $\omega \sim \Delta$, we then have 
 $b_\omega\sim 1$ {\it independently}
 on how large $b$ is. 
This in turn implies that at these frequencies, the system crosses 
over into the
Fermi liquid regime, and the behavior of $b_\omega$ at $\omega \leq \Delta$ is
qualitatively the same as for small $b$.
 Clearly then, 
$b_\omega$ should drop at around $\omega=2\Delta$ no matter how large $b$ is.
Our numerical solution of (\ref{inteq}) which we present
in Fig.~\ref{omsf} confirms our analytical result. We see that while the high
frequency part of $b_\omega$
evolves with increasing $b$
from frequency independent to a linear in frequency behavior, a
sharp drop in $b_\omega$ survives even for large $b$. 

\begin{figure} [t]
\begin{center}
\leavevmode
\epsfxsize=7.5cm
\epsffile{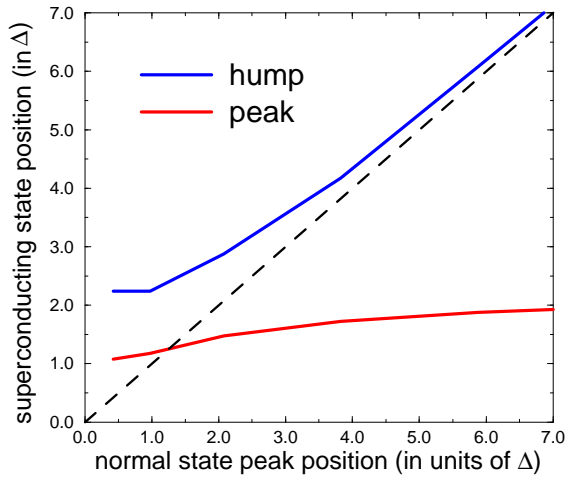}
\end{center}
\caption{The frequency position of the quasiparticle peak and the hump in the
superconducting state
versus position of the hump in the normal state obtained from
Fig.~\protect\ref{sftot}.}
\label{ns_vs_sc}
\end{figure}

The strong reduction of $b_\omega$
at low frequencies yields a crossover in the 
system behavior from a non-Fermi liquid, strong
coupling behavior at $\omega > 2 {\Delta}$ to a conventional
Fermi-liquid behavior at $\omega < 2 {\Delta}$. In the latter case, 
the spectral function $A({\bf k}_F,\omega)$ indeed possesses
 a sharp quasiparticle peak at $\omega = {\Delta}$ and 
rapidly decreases at higher frequencies.
 Combining these two limits, we find three distinct features in 
$A({\bf k}_F,\omega)$:
 a quasiparticle peak at $\omega \sim {\Delta}$, a dip at
$\omega \sim 2{\Delta}$ and a hump at 
$\omega =\Delta b/4$  which at $b \gg 1$
is parametrically larger than the peak frequency 
(see Fig.~\ref{sftot}). These results are fully consistent with the 
 photoemission data~\cite{norman,shennat}. 
Furthermore, analyzing the form of the spectral function Eq.(\ref{scg}) for
various momenta, we find that the quasiparticle peak does not disperse with ${\bf k}$ as the region of Fermi liquid behavior does not extend farther than $2 \Delta$ away from the Fermi surface. Instead,
the peak gradually decreases in the amplitude as one moves away from 
${\bf k}_F$.
 This is clearly seen in Fig. \ref{sftot}b,c. 
In contrast, the position of the hump follows 
$\omega \propto (\Delta^2_k + \epsilon^2_k)$. 
Near $(0,\pi)$, where $\epsilon_{\bf k}$ is small, the 
dispersion is weak, whereas further away from the Fermi surface it 
disperses
with ${\bf k}$ and gradually recovers the normal state dispersion.
In Fig~\ref{ns_vs_sc}, we
 plotted the frequency position of the quasiparticle peak and hump in the superconducting state vs. the normal state position of the hump.  
This dependence on momentum is also fully consistent with the 
photoemission data~\cite{norman}.

\begin{figure} [t]
\begin{center}
\leavevmode
\epsfxsize=7.5cm
\epsffile{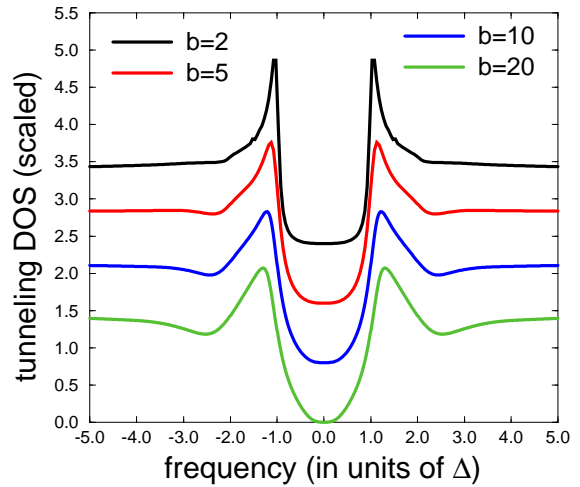}
\end{center}
\caption{The tunneling density of states in the superconducting state. 
Observe the development of a dip  with increasing $b$.}
\label{dos}
\end{figure}

Next we compute the 
density of states $N(\omega)$ in the 
superconducting state assuming that the
 dominant contribution to $N(\omega)$ 
comes from momenta near $(0,\pi)$. 
Integrating $A({\bf k},\omega)$ over $\epsilon_{\bf k}$, we obtain
\begin{equation}
N(\omega) \propto Re 
\frac{\Sigma (\omega)}{\sqrt{\Sigma^2(\omega) -\Delta^2}} \ .
\label{N}
\end{equation}                                           
 The plots of 
$N(\omega)$ for various $b$ are presented in  Fig.~\ref{dos}. 
We see that for all values of $b$, N$(\omega)$ possesses
 a peak at $\omega  = a\Delta$ where $a \approx 1$ for small $b$ 
and gradually increases with increasing $b$. For larger $b$, 
$N(\omega)$ displays a dip
 at frequencies  slightly larger than $2 \Delta$; the amplitude of the dip
increases
with $b$. Above the dip, $N(\omega)$ increases as $\sqrt{\omega}$ and 
eventually saturates. 
These results are in full agreement with the  
tunneling data in  Ref.~\cite{fisher} except 
for the experimentally observed anisotropy between $N(\omega)$ 
for positive and negative frequencies for which we do not have an explanation.

Furthermore,  we find that 
for large $b$ (i.e., for underdoped
cuprates), the region near $(0,\pi)$ yields a
substantial  contribution  to $N(\omega)$ even for $\omega <
{\Delta}$, which possibly overshadows the contribution from the 
nodes of $\Delta_{\bf k}$. Specifically, we found from Eq.(\ref{N}) that
$N(\omega) \propto \omega^3$ for very 
small frequencies which crosses over to $N(\omega) \propto \omega$
around $\omega \sim \Delta/2$. This behavior is very similar to
the one observed in underdoped cuprates. On the other hand, for small
 $b$ (i.e., for strongly overdoped cuprates), 
the contribution from the nodes is indeed relevant and
 our results underestimate $N(\omega)$ for $|\omega| \leq \Delta$. 

Finally, we discuss the value of the gap and its variation with doping.
Consider first the location of the quasiparticle peak in $A({\bf k},\omega)$. 
Our theory  predicts that at some distance away from the Fermi surface, it
should be located at $\omega \approx 2 {\Delta}$.
 Applying this result to
nearly optimally doped  Bi$2212$ materials studied  
in Ref.~\cite{norman,shennat}, 
we obtain 
${\Delta} \sim 25-30$ meV.
 Almost the same 
result is obtained
by extracting $2{\Delta}$ from the onset of the dip 
 in measured $N(\omega)$~\cite{fisher}. 
 Furthermore, for $T_c = 83K$
underdoped material, our analysis of the tunneling data yields 
almost the same value of ${\Delta}$ as at optimal doping. We therefore
argue that  ${\Delta}$  almost saturates around optimal doping, and
the observed increase of the peak frequency in $N(\omega)$ 
with decreasing doping is mostly due  
to strong coupling effects which shift the peaks towards higher frequencies.
Notice also that a comparison of $\Delta$ with $\omega_{sf} (\infty)$ extracted
from  NMR data \cite{pines} in the normal state confirms our assertion that 
$b>1$ already at optimal doping.
               
To summarize, in this paper we present the explanation of the unusual 
peak/dip/hump features observed in photoemission experiments on Bi$2212$
 at $T \ll T_c$. We argue that these features are explained by the
interaction of fermionic quasiparticles
with overdamped spin fluctuations. 
We show that the strong spin-fermion interaction combined with the 
feedback effect
on the spin damping due to fermionic pairing yield
 a Fermi-liquid form of the fermionic spectral function
at $\omega < 2 \Delta$ where
$\Delta$ is the maximum value of the superconducting gap,
and a non-Fermi-liquid form for $\omega > 2 \Delta$. In the Fermi-liquid
regime, the spectral function near ${\bf k}_F$ displays the quasiparticle peak at  $\omega \approx \Delta$; in the non-Fermi-liquid regime above $2\Delta$ 
it possesses a broad
maximum (hump) at $\omega \gg {\Delta}$. In between the two regimes,
 the spectral function has a dip at $\omega \sim 2 {\Delta}$. 
We argue that our theory also explains 
the tunneling data for the superconducting density of states.
We predict that the superconducting gap saturates around optimal doping, 
and that
the observed increase of the peak frequency 
in the tunneling density of states with decreasing doping 
is chiefly due to strong coupling effects.

It is our pleasure to thank G. Blumberg, 
 R. Joynt, M. Norman, D. Pines and J. Schmalian
 for useful conversations. The research was supported by NSF DMR-9629839 
(A.C.) and in part by STCS
through NSF DMR-9120000 (D.M.).

\end{document}